\documentclass{article}

\usepackage{arxiv}
\usepackage{appendix}
\usepackage[utf8]{inputenc} 
\usepackage[T1]{fontenc}    
\usepackage{hyperref}       
\usepackage{url}            
\usepackage{booktabs}       
\usepackage{amsfonts}       
\usepackage{nicefrac}       
\usepackage{microtype}      
\usepackage{lipsum}		
\usepackage{graphicx}
\usepackage{float}
\usepackage{todonotes}
\usepackage{stmaryrd}

\usepackage{dsfont}

\usepackage{amsmath,amssymb}
\usepackage{subcaption}
\usepackage{algorithm}

\title{Online Change Point Detection in Molecular Dynamics With Optical Random Features}

\author{{Amélie Chatelain}\\
	LightOn\\
	Paris, France \\
	\texttt{amelie@lighton.ai} \\
	\And
	{Giuseppe Luca Tommasone} \\
	LightOn\\
	Paris, France \\
	\texttt{luca@lighton.ai} 
	\And
	{Laurent Daudet} \\
	LightOn\\
	Paris, France \\
	\texttt{laurent@lighton.ai} \\
	\And
	{Iacopo Poli} \\
	LightOn\\
	Paris, France \\
	\texttt{iacopo@lighton.ai} \\
}

\renewcommand{\headeright}{}
\renewcommand{\undertitle}{Technical Report}

\begin{document}
\maketitle

\begin{abstract}
Proteins are made of atoms constantly fluctuating, but can occasionally undergo large-scale changes. Such transitions are of biological interest, linking the structure of a protein to its function with a cell. Atomic-level simulations, such as Molecular Dynamics (MD), are used to study these events. However, molecular dynamics simulations produce time series with multiple observables, while changes often only affect a few of them. Therefore, detecting conformational changes has proven to be challenging for most change-point detection algorithms. In this work, we focus on the identification of such events given many noisy observables. In particular, we show that the No-prior-Knowledge Exponential Weighted Moving Average (NEWMA) algorithm can be used along optical hardware to successfully identify these changes in real-time. Our method does not need to distinguish between the background of a protein and the protein itself. For larger simulations, it is faster than using traditional silicon hardware and has a lower memory footprint. This technique may enhance the sampling of the conformational space of molecules. It may also be used to detect change-points in other sequential data with a large number of features.
\end{abstract}


\section{Introduction}

Today, about $60\%$ of High-Performance Computing (HPC) workload performs computational chemistry and material sciences tasks. These undertakings open the door to the discovery of new materials, enhance our understanding of biological mechanisms, and potentially allow for the discovery of new drugs.

One popular and straightforward technique for performing such computations is Molecular Dynamics (MD). MD computations follow trajectories of atoms over an extended period of time thereby providing detailed physical information as well as quantitative data on the chemical system at hand. Developed in the early 1950s, the technique has since increased in systems complexity with the increase in computational power afforded by Moore’s law. Larger and larger molecular trajectories can be achieved over longer periods of time.

HPC and computational chemistry developments have been intertwined over the years. On a generic hardware level, HPC architectures have evolved tremendously over the past 60 years as an answer to larger system computations. In parallel, the discovery of new algorithms has allowed ever faster implementations of these simulations (eg. Fast Multipole Method \cite{rokhlin1985}). Eventually, specific elements of the computation pipeline have been required to speed up some aspects of these simulations (e.g. Gravity Pipe (GRAPE) \cite{makino1998} or more recently Anton \cite{anton2008}). While the bulk of the computation is focused on obtaining trajectories, a real insight can only come through the analysis of said trajectories which in and of itself may require large resources. In particular, making sure that the different states of a protein have been explored is a significant task.

Biomolecular systems such as viruses are made of atoms constantly moving. They can take different shapes called conformations. Identifying transitions from one conformation to another one is of biological interest. Indeed conformational changes link the structure of a biomolecule to its function within an organism. Thus, knowing these transitions is key in the design of new drugs.

As the number of atoms studied has grown bigger, larger data streams are being produced at every timestep of these simulations. As a result, conformational changes can be tricky to identify, especially for large molecular systems, as changes can sometimes be very local. Processing and memory limitations are also in the way of understanding these transitions as they require full trajectories to be post-processed. Detecting these changes is expensive and performed by batch thereby requiring many computational resources. 

Machine learning techniques have been adapted to MD for a more agile, online process. Notably, Trstanova, Leimkuhler, and Lelièvre \cite{Trstanova2019} implemented  Diffusion Maps (DMaps), a dimensionality reduction technique, to enhance the exploration of the conformational spaces of molecules in MD. However, the use of this algorithm may not be the most effective way to identify conformational changes, compared to the use of techniques designed for similar tasks.

Change point detection focuses on detecting abrupt changes in time series, with applications in meteorology, finance, medicine, and entertainment. These methods may be limited by computational power as well as the dimensionality of the data considered. Keriven, Garreau and Poli \cite{Keriven2018} introduced the No-prior-knowledge Exponential Weighted Moving Average (NEWMA) algorithm, a model-free online detection algorithm which appears suited to the analysis of MD trajectories.

\paragraph{Contributions and outline} In this paper, we focus on using NEWMA and optical random features to analyse MD trajectories, and propose a strategy coupling NEWMA and DMaps to make analysing MD simulations more effective. We start by introducing the problem from a MD point of view and the value of DMaps in Section \ref{sec:md}. We then explore NEWMA as a method to study conformational changes and propose to combine DMaps and NEWMA in Section \ref{sec:onlinecp}. Section \ref{sec:results} compares the numerical results obtained using NEWMA to various methods. Additionally, we argue on the use of optical random features over silicon hardware. We present our conclusion remarks in Section \ref{sec:conclusion}.

We encourage the reader of this article to explore our publicly-available code used at \url{https://github.com/lightonai/newma-md}. 

\section{Diffusion Maps to Identify Conformational Changes}
\label{sec:md}

In this section, we introduce some of the key challenges of MD simulations, and detail how DMaps can be used to tackle these issues. 

\subsection{Conformational Changes: Sampling in Molecular Dynamics}

MD simulates the natural behavior of a system to discover physical properties of atoms and molecules. Its goal is to explore the conformational space of large molecules, i.e. all its possible structures. The challenge is sampling the free energy landscape of our molecule that varies with changes in conformations. It informs on the equilibrium thermodynamics and dynamical properties of the system, as well as offering an atomic-scale insight. 

\begin{figure}[htpb]
    \centering
    \includegraphics[width=.3\textwidth]{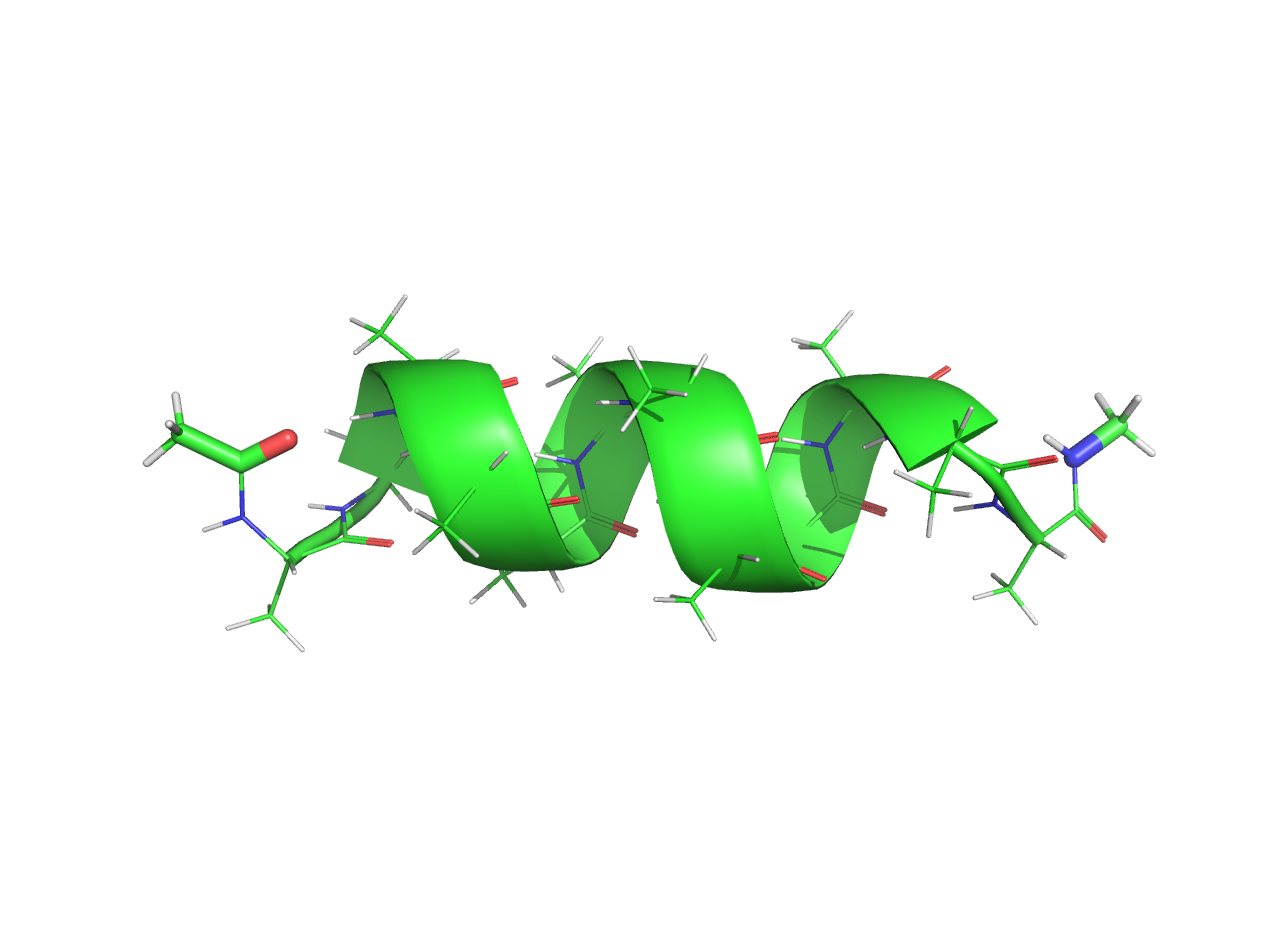}%
    \includegraphics[width=.3\textwidth]{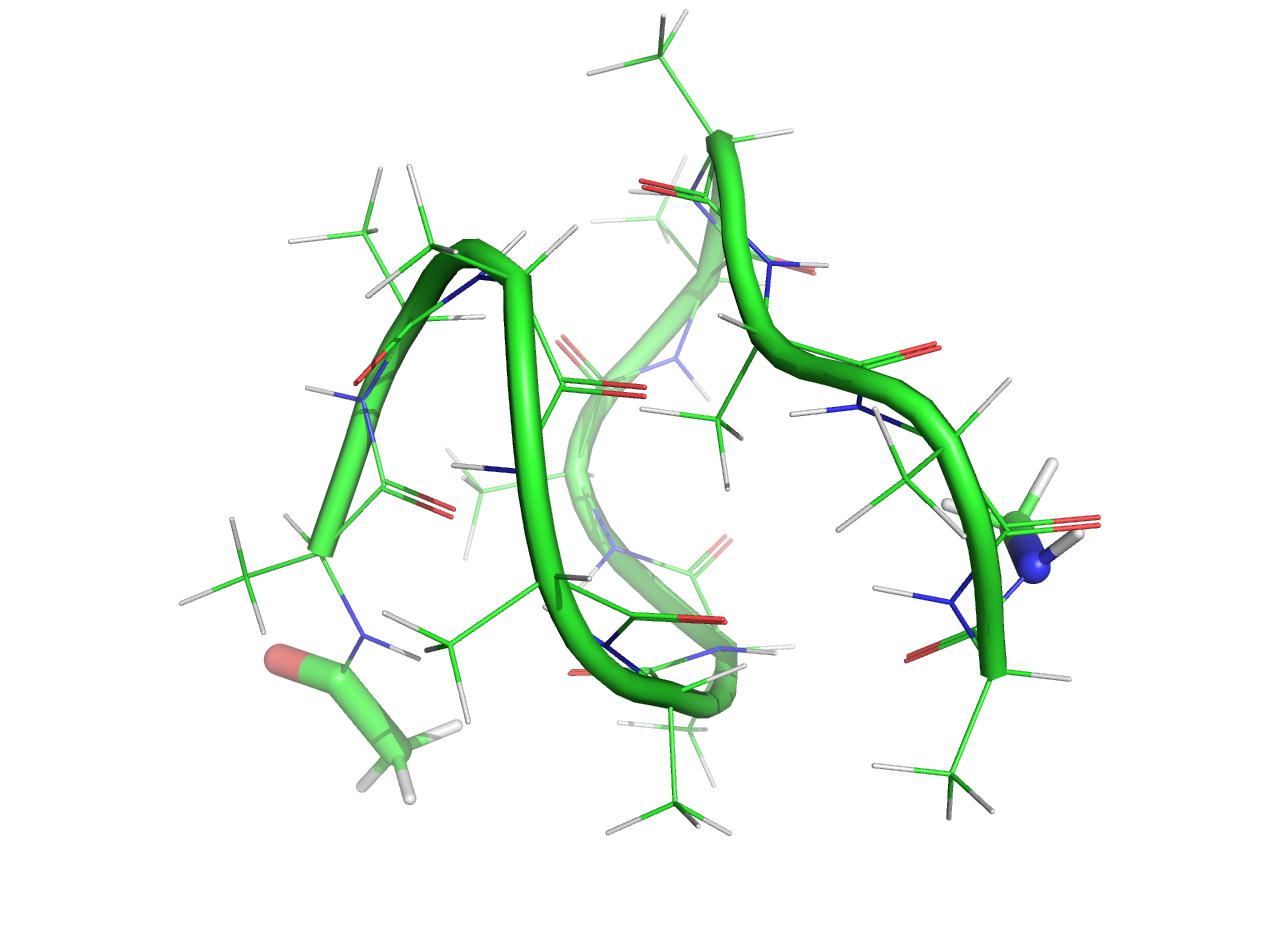}%
    \includegraphics[width=.3\textwidth]{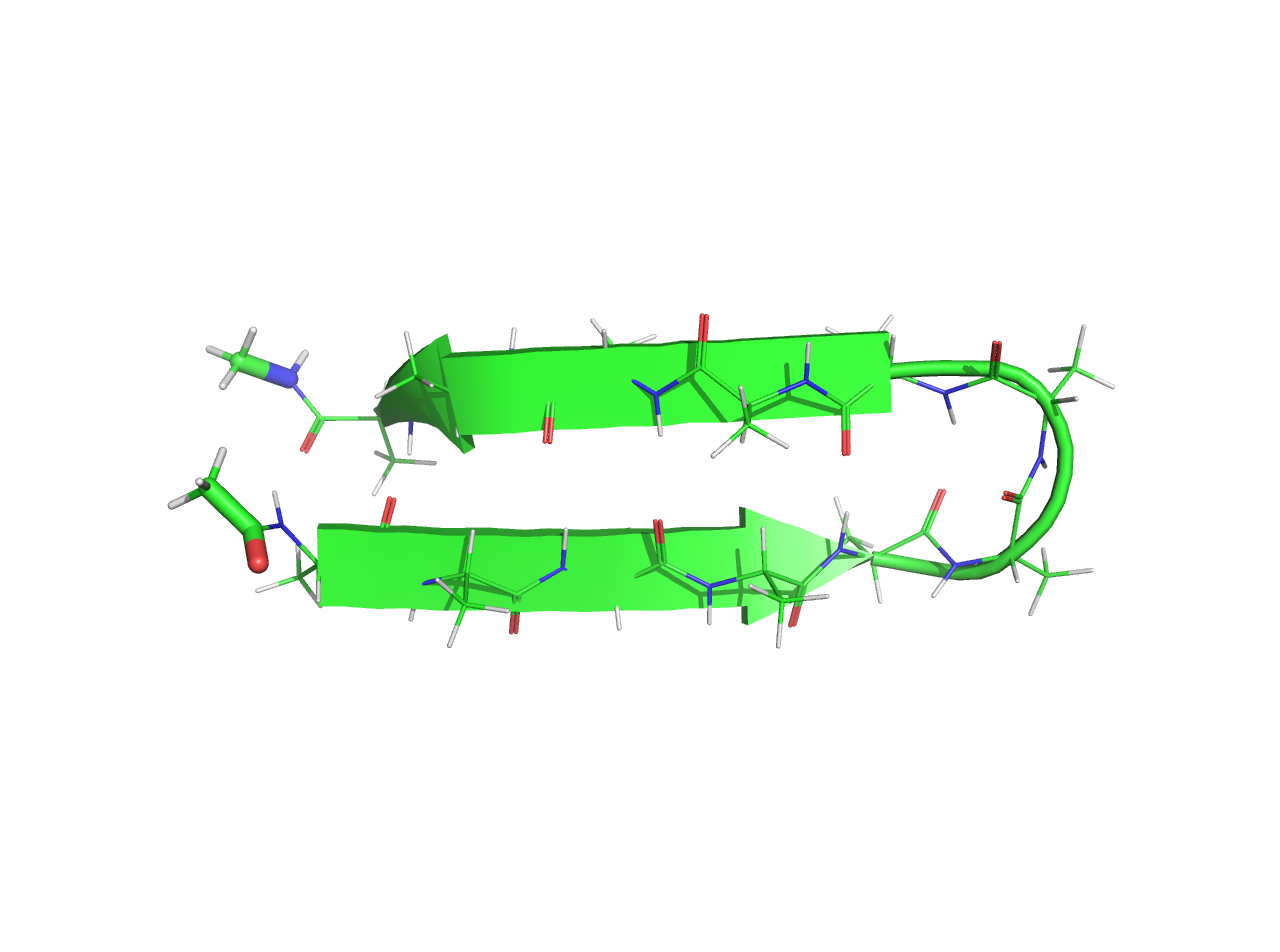}
    \caption{Different conformations of the deca-alanine protein (from Ref. \cite{Trstanova2019}).}
    \label{fig:decaalanine}
\end{figure}

Typically, MD simulations output time frames $F_t \in \mathbb{R}^{3n}$ at different timesteps $t \in \mathbb{N}$. These frames are vectors storing the three-dimensional coordinates of the $n$ atoms of a molecular system. However, there are still two main factors limiting the reliability of these methods. 

The first is known as the \textit{force field problem}: it is difficult to know accurately the potential energy function associated with the forces applied to the system. 

The second is the \textit{length of the simulations}: the timestep $\Delta t$ used in integration solvers is limited by the highest frequency vibration of the system. More precisely, MD simulations solve Newton’s laws of motion for a group of many atoms. Since they all move around constantly, their bonds fluctuate and some can do so at high frequency. In most cases, the Carbon-Hydrogen bond stretching puts an upper bound on that frequency thereby requiring integration steps $\Delta t$’s to be of the order of the femtosecond.

In the lower frequency range, the different conformations of a protein corresponding to local minima (see for example Figure \ref{fig:decaalanine}) --- or \textit{metastable states} --- are separated by high-energy barriers. As a result, transitions between such structures are rare events that occur on a timescale of the microsecond and sometimes up to the millisecond. In order to see a single change in the free energy landscape, we potentially need to perform a sequential integration of Newton’s equations more than a billion times.

To counter this issue, enhanced sampling methods, such as metadynamics \cite{Laio2008}, have been developed in theoretical chemistry. The idea behind such strategies is to modify the potential energy of a molecule by adding an extra term called \textit{bias}. This bias will decrease the energy barrier between conformations, forcing the system out of the metastable state where it is located. 

\begin{figure}
    \centering
    \includegraphics[width=.6\textwidth]{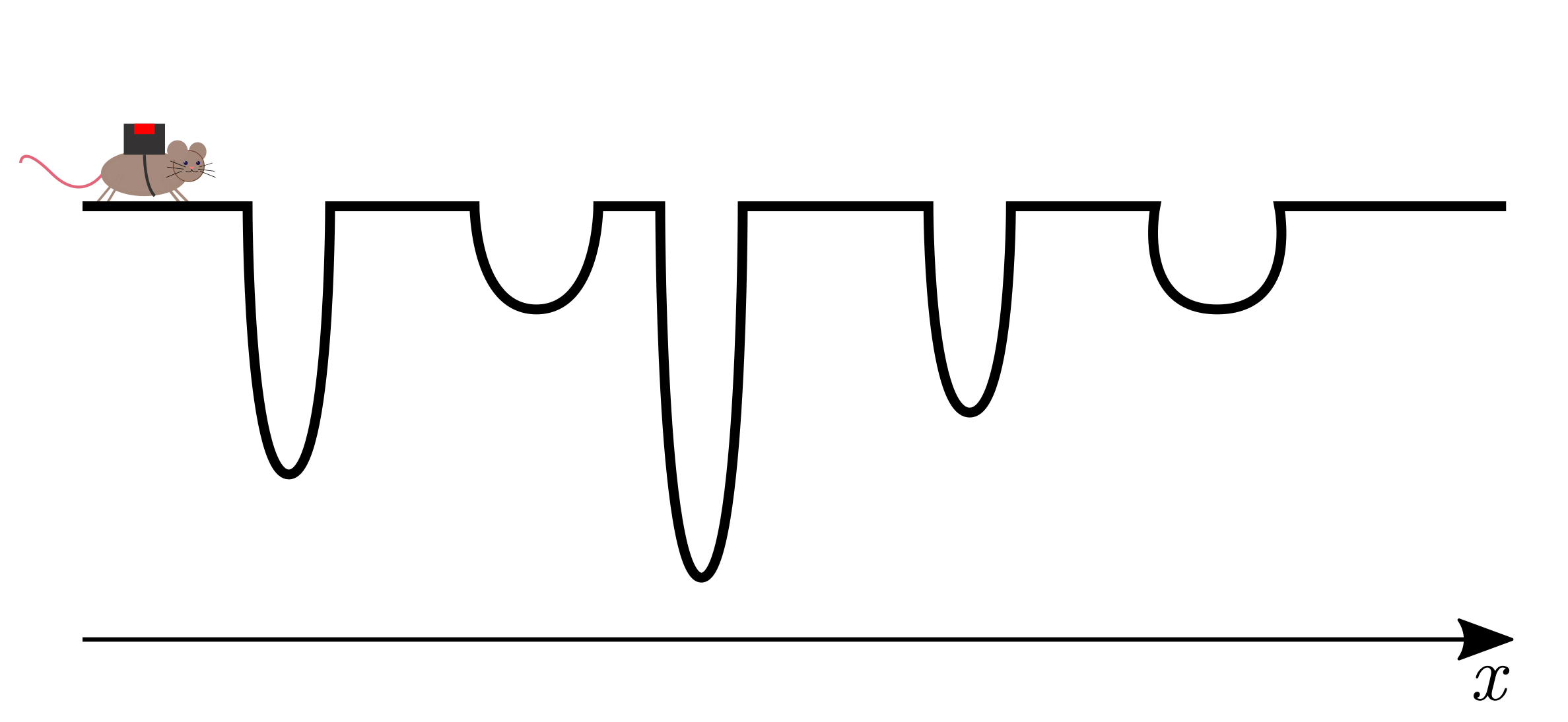}
    \includegraphics[width=.6\textwidth]{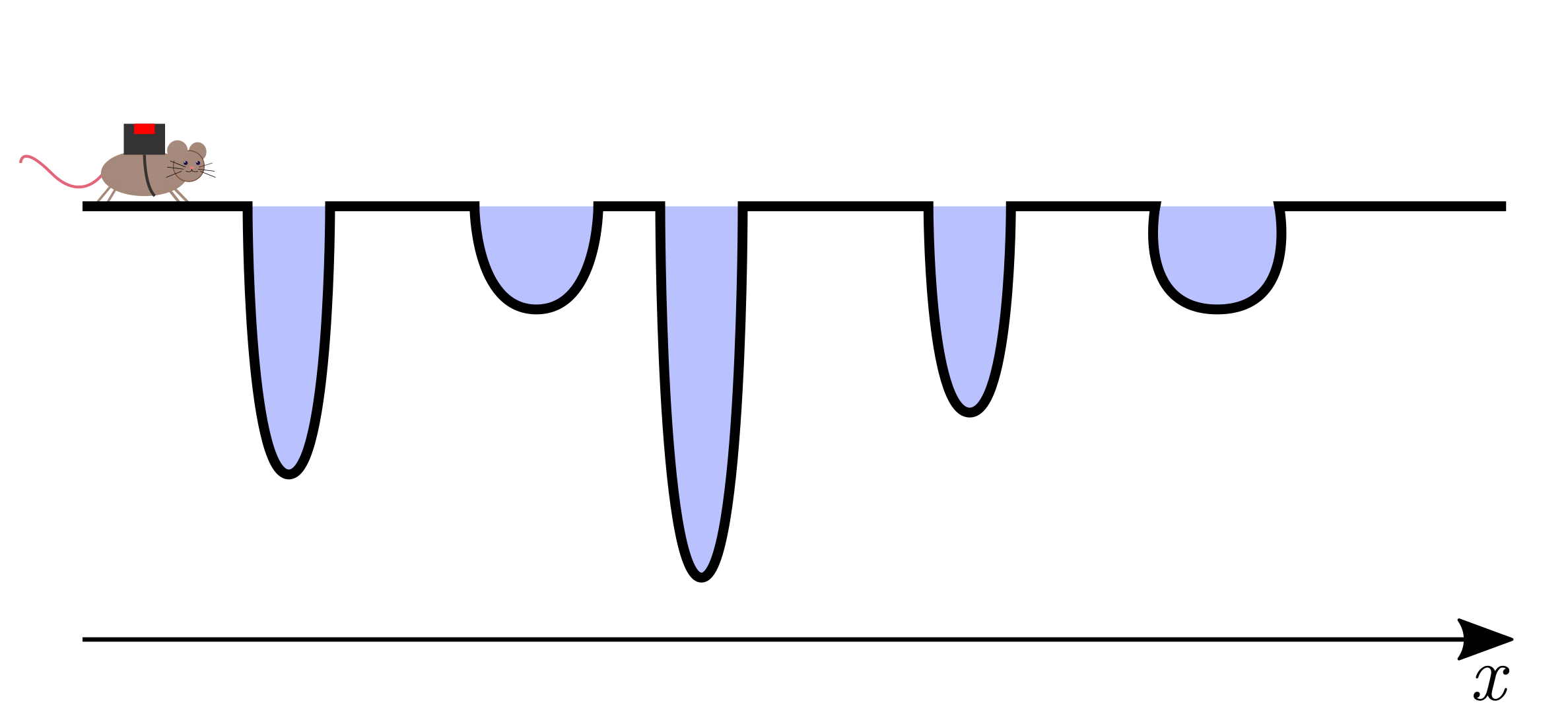}
    \caption{R\'emy the rat in his holey world, before exploration (top) and after (bottom). We pour water in the holes to help him get out of them whenever he is stuck.}
    \label{fig:remy}
\end{figure}

We introduce R\'emy the Rat to illustrate this idea. R\'emy walks in a land filled with holes, and we would like to be able to map those holes and know their depth (that is, we want to know all possible shapes of a protein and associated energies). Now, imagine that R\'emy’s land is in a tank, and we can only see it from a bird’s-eye view. That makes it very difficult to see the holes or make any height measurements.

Our solution is to attach a GPS tracker to R\'emy’s back --- see Figure \ref{fig:remy}. As R\'emy walks around, we can track his position on the $x$-axis. Then, we simply say that if his position does not vary much for a given time, that means that he fell in a hole. Assuming that R\'emy is a strong swimmer, to set him free, we pour water at his location, until we see that he is able to move freely once again. By keeping track of the amount of water we have poured, we can know how deep the hole was. After a while, R\'emy will be able to move across his entire land: we have explored all the conformations of our molecule. This is the idea behind techniques such as metadynamics.

\begin{figure}[htpb]
\centering
    \begin{subfigure}[b]{0.35\textwidth}
    	\includegraphics[trim={.8cm .8cm .8cm .8cm}, clip, width=\textwidth]{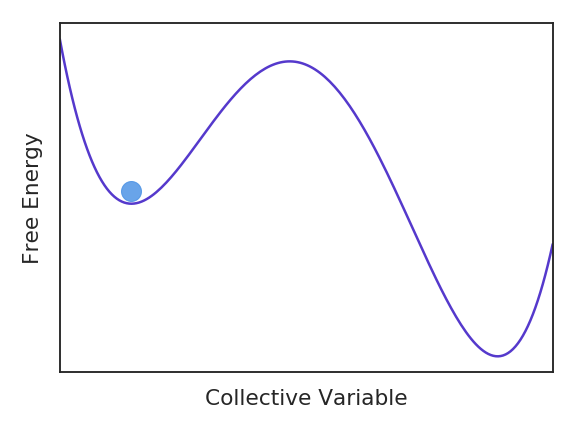}
    	\caption{}
    	\label{fig:metadym1}
    \end{subfigure}%
    \begin{subfigure}[b]{0.35\textwidth}
    	\includegraphics[trim={.8cm .8cm .8cm .8cm}, clip, width=\textwidth]{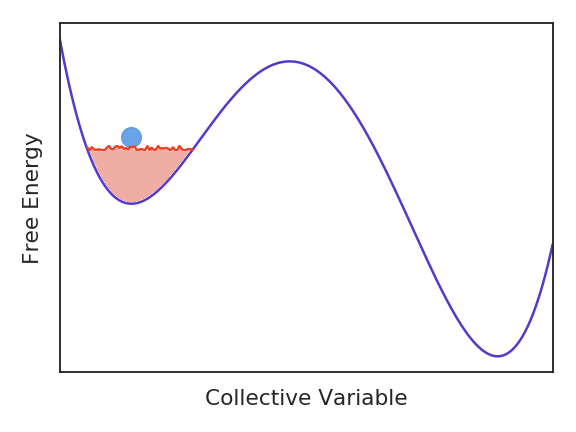}
    	\caption{}
    	\label{fig:metadym2}
    \end{subfigure}
    
    \begin{subfigure}[b]{0.35\textwidth}
    	\includegraphics[trim={.8cm .8cm .8cm .8cm}, clip, width=\textwidth]{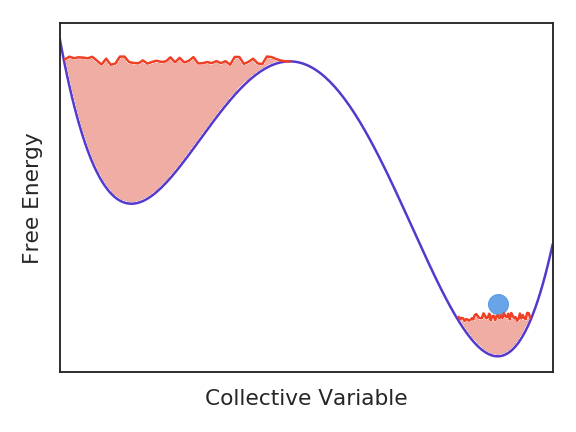}
    	\caption{}
    	\label{fig:metadym3}
    \end{subfigure}%
    \begin{subfigure}[b]{0.35\textwidth}
    	\includegraphics[trim={.8cm .8cm .8cm .8cm}, clip, width=\textwidth]{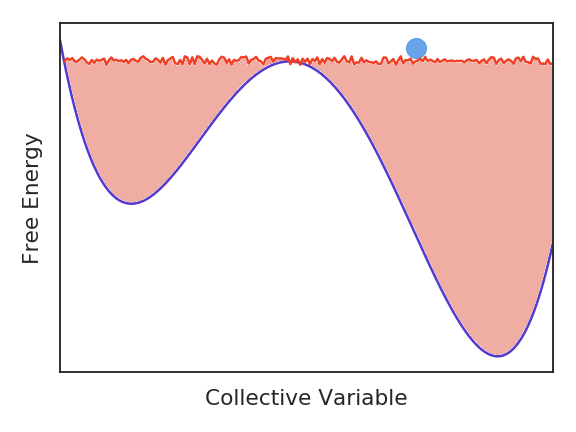}
    	\caption{}
    	\label{fig:metadym4}
    \end{subfigure}
    
    \caption{Schematic representation of enhanced sampling strategies based on CVs. All of these figures represent the free energy of the system (solid blue line) as a function of one CV. At first (Fig. \ref{fig:metadym1}), the system (dark blue bullet) is trapped in a local minimum. As a high-energy barrier separates this minimum from the global minimum, the simulation is only able to sample the region around the local minimum. Therefore, the bias (red line) is only added in that region - see Fig. \ref{fig:metadym2}. As the reference walker spends more time trapped in the local minimum, the bias increases until it becomes large enough for the system to reach the global minimum (Fig. \ref{fig:metadym3}), in which a smaller bias starts to build up. Finally, the bias is larger than the free energy barrier, and the simulation is able to sample easily the entire energy landscape. The free energy can be recovered by inverting the bias. }
    \label{fig:metadym}
\end{figure}

Such algorithms make use of collective variables (CVs) that are one-dimensional coordinates assumed to describe the system in its current state. These CVs enable the “guiding” of the simulation as shown in the procedure featured in Figure \ref{fig:metadym}. By adding a bias, we can push a system out of its “prison” through the addition of an extra force.

Typically, CVs can be physical coordinates such as the angle between certain bonds in molecules. In that case, they can be chosen using intuition. Unfortunately, it is not always that simple, especially for very large systems. Several methods have been developed to automatically detect such coordinates. In particular, Trstanova, Leimkuhler, and Lelièvre \cite{Trstanova2019} proposed a method based on a dimensionality reduction algorithm called DMaps. We explore the use of this technique in the next section.

\subsection{Diffusion Maps and Collective Variables}
\label{sec:dmaps}

This subsection focuses on the DMaps algorithm, a nonlinear dimensionality reduction technique, and its application to MD results. Trstanova, Leimkuhler, and Lelièvre \cite{Trstanova2019} have shown that this method can be used to identify when the system has reached a metastable state, as well as CVs in that state. We start by discussing DMaps, and then apply them to MD simulations. 

\subsubsection{Diffusion Maps}

Let $\textbf{x} = \left\lbrace x_1,\, x_2,\, ...\, x_m \right\rbrace$ be a set of $m$ vectors, with $x_i \in \mathbb{R}^N$ for $i \in \llbracket 1,\, m \rrbracket$. Our objective is to find a nonlinear mapping into a $k$-dimensional space, with $k<N$, while preserving the structure of the dataset. The original DMaps algorithm, introduced in Ref. \cite{coifman2005}, is described in Algorithm \ref{alg:DM}. 

\begin{algorithm}
\caption{Diffusion map algorithm.}
\label{alg:DM}

\begin{enumerate}
	\item Choose an isotropic diffusion kernel, such as the RBF kernel
	\begin{equation}
		k \left( x_i, x_j \right) = \exp(-\gamma\Vert x_i - x_j \Vert^2),
		\label{eq:rbfkernel}
	\end{equation}
with $\gamma \in \mathbb{R}^{+*}$ an hyperparameter, and define the $m \times m$ \textit{diffusion matrix} $L$ such that $L_{i,j} = k \left( x_i, x_j \right)$. The diffusion matrix is positive semi-definite: it has non-negative, real eigenvalues.
	
	\item Define the diagonal matrix $D$ such that $D_{i, i} = \sum_{j=1}^m L_{i, j}$, and normalize $L$ according to the hyperparameter $\alpha \in \left[ 0, 1 \right] $
\begin{equation}
L^{\left( \alpha \right)} = D^{-\alpha} L D^{-\alpha}.
\end{equation}
	
	\item Define the diagonal matrix $D^{\left( \alpha \right)}$ such that $D^{\left( \alpha \right)}_{i, i} = \sum_{j=1}^m L^{\left( \alpha \right)}_{i, j}$, and form the normalized matrix $M$
	\begin{equation}
	M = \left( D^{\left( \alpha \right)} \right)^{-1} L^{\left( \alpha \right)}.
	\end{equation}
$M$ defines the transition matrix of a Markov process. Taking larger and larger powers of this matrix, $M^t$ with $t \in \mathbb{N}^*$, reveals the geometric structure of the studied data set at larger and larger scales. The probability to transition from one state to another in $t$ steps is given by $M^t$. This corresponds to a diffusion process. 
	 \item Compute the $k$ largest eigenvalues $\lambda_i$, $i=1, ...k$, and corresponding eigenvectors $\psi_i$, $i=1, ...k$, and build the embedding
	 \begin{equation}
		 \Psi_t \left( \textbf{x} \right)= \left( \lambda_1^t\psi_1\left( \textbf{x} \right),\lambda_2^t\psi_2 \left( \textbf{x} \right) ,...,\lambda_k^t\psi_k \left( \textbf{x} \right)\right).
	 \end{equation}
The data points can now be represented using their \textit{Diffusion Coordinates} (DCs) $\lambda_i \psi_i$, $i=1,...k$. The eigenvector corresponding to the largest eigenvalue, $\lambda_1 = 1$, is always a constant vector, $\psi_1 = \mathds{1}$. Therefore, it does not contribute to the spectrum of the diffusion matrix. Excluding $i=1$, we call \textit{dominant} the coordinates $\lambda_i \psi_i$ with a large enough associated eigenvalue.
\end{enumerate}
\end{algorithm}

The hyperparameters $\gamma$, $\alpha$, $t$ should be selected to generate a low-dimensional embedding that preserves the main features of the data set. More details are discussed in the Appendix. Note that the dimensionality of the data can be reduced with a different technique prior to this algorithm, such as \textit{Random Projections} (RPs) \cite{bingham2001}.

In the Appendix, we explore the use of DMaps on the swiss roll data set \cite{marsland2009} in Section \ref{sec:swissroll} and we discuss the choice of hyperparameters for the algorithm in Section \ref{sec:hyperparam}. We detail now how DMaps can be used on data produced by MD simulations to enhance sampling.

\subsubsection{Applying Diffusion Maps to Molecular Systems}

The authors of \cite{Trstanova2019} proposed an algorithm to enhance the sampling of the configuration space of a given molecule using DMaps locally. The interest in using DMaps to MD trajectories is two-fold. 

Firstly, by computing the spectrum of the diffusion matrix, it is possible to detect metastable states as a collection of states for which the eigenvalues are almost stationary. Thus, a change in the spectrum enables us to detect a conformational change.

Secondly, using Pearson's correlation coefficients \eqref{eq:pearson} between the DCs and the physical coordinates of the problem, it is possible to identify (physical) CVs. Those correspond to the slowest modes of the metastable state. This provides important information to enable the system to escape the metastable state, enhancing the sampling. Note that using CVs is more interesting than using diffusion coordinates, as they correspond to physical quantities (coordinates, angles, etc.). Therefore, they are defined on all the configuration space, while diffusion coordinates are only defined locally.

\begin{figure}
    \centering
    \includegraphics[width=.8\textwidth]{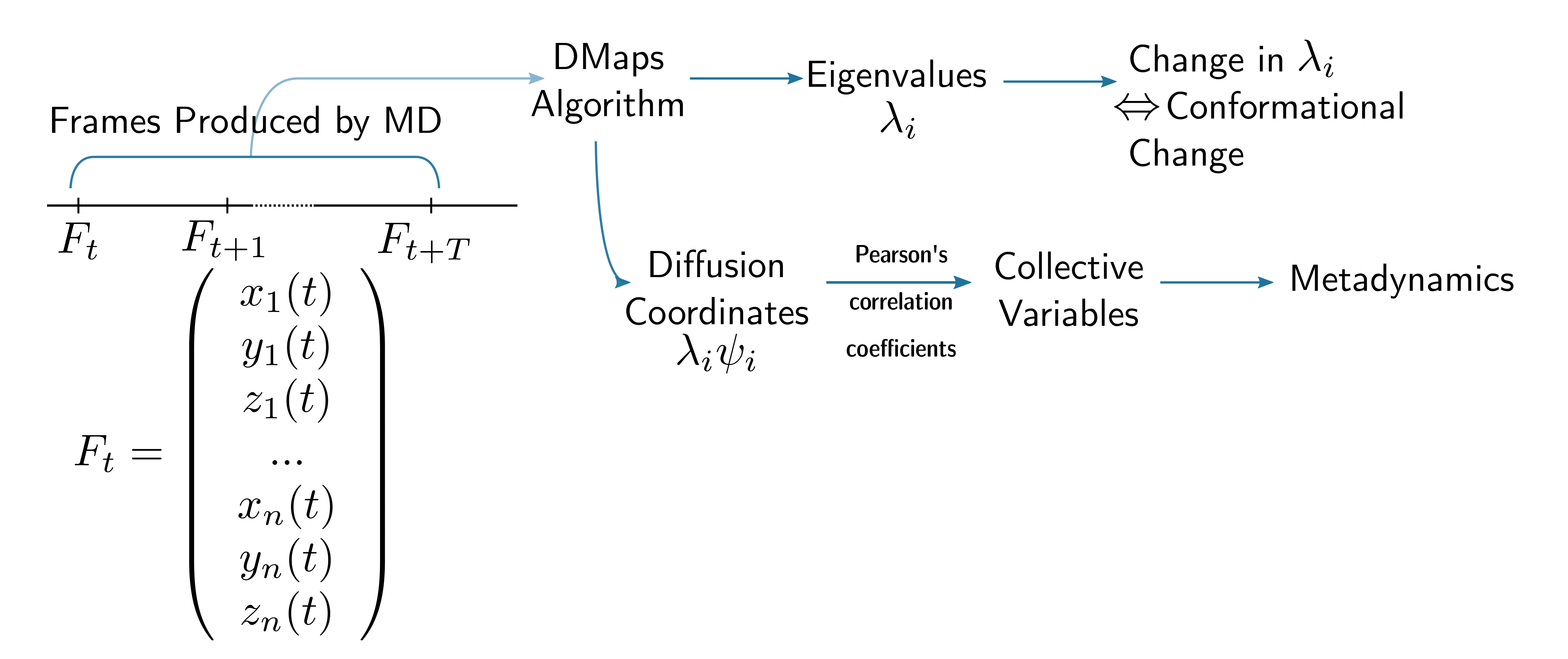}
    \caption{Outline of the use of DMaps for MD problems, as proposed in Ref. \cite{Trstanova2019}. $F_t$ denotes the time frame produced at time $t$, which is a vector storing the coordinates of all $n$ atoms.}
    \label{fig:mdprocedure}
\end{figure}

 Algorithm \ref{alg:enhancedsamp} outlines such a procedure. In this method, DMaps are used in steps 1 and 4 --- computing the spectrum of the diffusion matrix to detect the entrance and exit of a metastable state --- and in step 2 --- computing the diffusion coordinates to identify the CVs. Figure \ref{fig:mdprocedure} illustrates the steps of the method.

\begin{algorithm}[htpb]
\begin{enumerate}
\item Run the MD, computing the spectrum of the diffusion matrix every $T \in \mathbb{N}^*$ steps, until it is almost stationary. This means that the system has reached a metastable state. 
\item Identify the CVs as the physical coordinates which are the most correlated to the diffusion coordinates. 
\item Using the learnt CVs, build a bias in the metastable state. 
\item A change in the spectrum means that the system has left the metastable state it was exploring. Restart from step 1, keeping the bias built before. 
\end{enumerate}
\caption{Enhanced sampling of the configuration space based on DMaps, as proposed in Ref. \cite{Trstanova2019}.}
\label{alg:enhancedsamp}
\end{algorithm}

The use of such a procedure relies on applying the DMaps algorithm every $T$ steps. As we see in the next section \ref{sec:newma}, this is far from optimal. Hence, we explore alternatives.

\section{Online Change-Point Detection to Identify Conformational Changes}
\label{sec:onlinecp}

In this section, we explore the limitations of the application of DMaps alone to MD problems, and offer the use of NEWMA as an addition to the strategy for enhanced sampling introduced before.

\subsection{Using Diffusion Maps to Detect Conformational Changes: Limitations}

Previously, we have highlighted how DMaps can be used to identify metastable states. Indeed, computing the spectrum of the diffusion matrix, we can identify a collection of states with almost-stationary eigenvalues as a metastable state, while sharp changes characterize an exit of this state. However, this method has some limitations.

The first one is that it forces us to treat several time frames as \textit{batches} of $T$ points on which to apply the DMaps algorithm. Too small batches are rather unpractical, as the eigenvalues may then vary too much, while too large batches are more demanding numerically.

In addition to that, this means that batches of $T$ time frames have to be stored, which can lead to a significant \textit{memory footprint} for as the number of atoms $n$ grows, since the frames are vectors of dimension $n$.

Note that the DMaps algorithm requires to calculate pair-wise distances between the $T$ frames, re-normalizing and extracting the eigenvalues from the diffusion matrix. The \textit{computational cost} of these operations is quite significant. 

Furthermore, as discussed in Paragraph \ref{sec:hyperparam}, the hyperparameters of the DMaps procedure can be somewhat tricky to find, and come with no guarantees to be valid from one MD trajectory to the next. 

Finally, to determine if the eigenvalues are almost stationary, one needs to quantify the way they are changing. This involves defining a threshold, which may again not be universal. 

In the rest of this article, we explore how the use of NEWMA, an online change-point detection algorithm, can be used in combination with the Algorithm \ref{alg:enhancedsamp} to enhance sampling in MD. 

\subsection{NEWMA: A Method For Online Change-Point Detection}
\label{sec:newma}
Online change-point detection aims at detecting and flagging sharp changes in a stream of data points. The challenge is to identify such transitions as soon as they occur and limit the number of false alarms. Online methods do not require to store raw data points, hence having a lower memory footprint as well as being valuable when the data is sensitive. While the classical Exponentially-Weighted Moving Average (EWMA) estimation meets those criteria, it relies on prior knowledge of the data. 

\begin{algorithm}[htpb]
\caption{NEWMA, as proposed in \cite{Keriven2018}, applied to a MD problem.}
\label{alg:nemwalgo}
Let $F_t \in \mathbb{R}^{3n}$ be a vector containing the coordinates of  $n$ atoms at time $t$, $\Psi : \mathbb{R}^{3n} \rightarrow \mathbb{R}^{k}$ a random embedding, with $k \in \mathbb{N}^*$, $0<\lambda <\Lambda <1$ two forgetting factors, $\tau \in \mathbb{R}^{+*}$ a threshold and $z_0 = z_0'\in  \mathbb{R}^{k}$ two initial values. 
For $t \in \mathbb{N}^{*}$
\begin{enumerate}
\item Compute 
\begin{align*}
z_t = \left( 1 - \Lambda \right) z_{t-1}& + \Lambda \Psi \left( F_t \right), \\
z'_t = \left( 1 - \lambda \right) z'_{t-1}& + \lambda \Psi \left( F_t \right).
\end{align*}
\item If $\Vert z_t - z'_t \Vert \geq \tau$, flag $t$ as a change-point. 
\item Increment $t=t+1$ and restart from step 1.
\end{enumerate}
\end{algorithm}

The idea behind NEWMA is to compute a detection statistic as the difference between two EWMA statistics. This quantity is then compared to an adaptive threshold. If this difference is above said threshold, the algorithm flags that point as a change-point (Algorithm \ref{alg:nemwalgo}). This method has the advantage of requiring no prior knowledge about the change-points or the trajectory studied, and can be computed online. Furthermore, the hyperparameters of the algorithms are selected by a heuristic computed only on the window size — that is, the number of recent samples that are to be compared with older ones. More details about the algorithm as well as the heuristics for its hyperparameters search can be found in Ref. \cite{Keriven2018}.

The detection statistics are calculated using RPs as the embedding $\Psi$. They can be computed on CPU using methods such as Random Fourier Features (NEWMA RFF on CPU) \cite{rahimi2008} or FastFood (NEWMA FF on CPU) \cite{le2013}. Recently, an Optical Processing Unit (OPU) has been developed at LightOn \cite{lighton2020}, computing random features through light interactions with matter, at a constant computational cost and no memory expense. We use it to compute the detection statistics (NEWMA RP on OPU).

\subsection{Proposed Strategy: NEWMA and Diffusion Maps}

We now explore the possibility to combine DMaps with the work of Ref. \cite{Keriven2018}, which provides a low-memory-footprint method to detect change-points with an adaptive threshold scheme. 

\begin{figure}
    \centering
    \includegraphics[width=\textwidth]{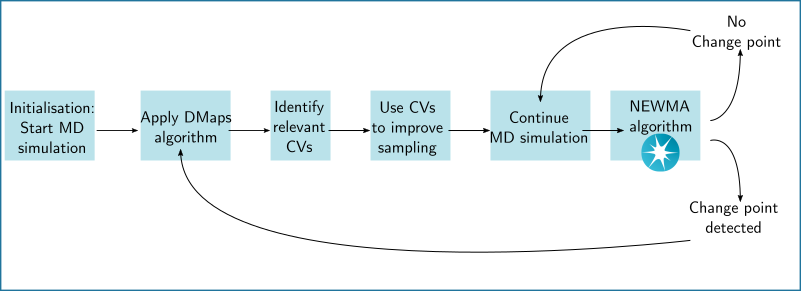}
    \caption{Strategy to enhance sampling in MD, using the NEWMA algorithm to detect conformational changes.}
    \label{fig:enhancednewmasamp}
\end{figure}

Our idea is to use such a procedure to identify conformational changes, replacing the steps $1$ and $4$ of Algorithm \ref{alg:enhancedsamp}. This allows us to apply the DMaps algorithm only in steps $2$, to identify the CVs through the diffusion coordinates, and use them to enhance the sampling. Figure \ref{fig:enhancednewmasamp} summarizes this method. In the next subsection, we show numerical results and discuss the performances required. 

To go back to our analogy with R\'emy the Rat, saying that R\'emy is stuck because he hasn’t moved that much lately presents some issues:
\begin{itemize}
    \item  What does “moving that much” mean? In practice, that means defining a threshold. Yet, we do not want our mapping to be dependent on a user-defined quantity.
    \item This method implies that we are monitoring R\'emy’s position at all times, which means a lot of resources: a human watching, a satellite constellation, etc. Correspondingly, sampling methods and DMaps in MD require lots of computational resources. Not only trajectories have to be computed, but then they need to be analyzed online to determine if the molecule is stuck in a conformation or not.
\end{itemize}

The use of an accelerometer circumvents those issues: accelerometers do not use satellites as GPS trackers do and are much simpler and cheaper. Using one, we can detect if R\'emy is suddenly falling, and we know that this means he is falling into a hole. We can save some time and some energy. We do not have to track R\'emy at all times, only to put an alarm when the accelerometer says that R\'emy fell. This is essentially what using NEWMA does for our MD simulations. 

NEWMA makes it possible to use the DMaps algorithm only once a conformational change has been detected, instead of using it every $T$ batch of time frames. Once a change-point has been identified and only then, $T$ points are stored to be used in DMaps and learn CVs. Therefore, this strategy has a much lower computational and memory cost than the one proposed in Algorithm \ref{alg:enhancedsamp}. In addition, it also solves our issues of being model dependent. 

In summary, in the new MD calculation pipeline shown in Figure \ref{fig:enhancednewmasamp}, NEWMA-based detection can be efficiently computed at every timestep while the coarser-grained, low-resolution DMaps algorithm can be used to obtain the coordinate representation in the manifold dataset when it is needed. This approach leads the way to faster global sampling studies, such as metadynamics. We present our numerical experiments in the next section. 

\section{Results}
\label{sec:results}

In this section, we focus on our numerical results, testing NEWMA as a method to detect conformational changes in MD, and compare it to DMaps. We also compare the different methods to compute RPs discussed in Section \ref{sec:newma}. All of our experiments were run on the LightOn Cloud \cite{lighton2020}, using an Intel Xeon Gold $6128$ $3.40$GHz CPU and a LightOn Aurora OPU. The code is available at \href{https://github.com/lightonai/newma-md}{\texttt{https://github.com/lightonai/newma-md}}.

\subsection{Comparison With Other Methods: DMaps}

On March 11, 2020, the World Health Organization (WHO) has declared the ongoing coronavirus disease COVID-19 a pandemic. This infection is caused by the virus Severe Acute Respiratory Syndrome Coronavirus 2 (SARS-CoV-2). To face this challenge, “the greatest since World War Two” according to Angela Merkel, a significant portion of the scientific community has been devoted to the research on the virus. In particular, simulations of the SARS-CoV-2 virus have been conducted to determine the efficiency of certain drugs such as Lopinavir \cite{Cespugli2020}. This simulation includes $30\ 000$ time frames for $n=4\ 776$ atoms, that we use to showcase the efficiency of NEWMA.

\begin{figure}[htpb]
\centering
\includegraphics[width=.6\textwidth]{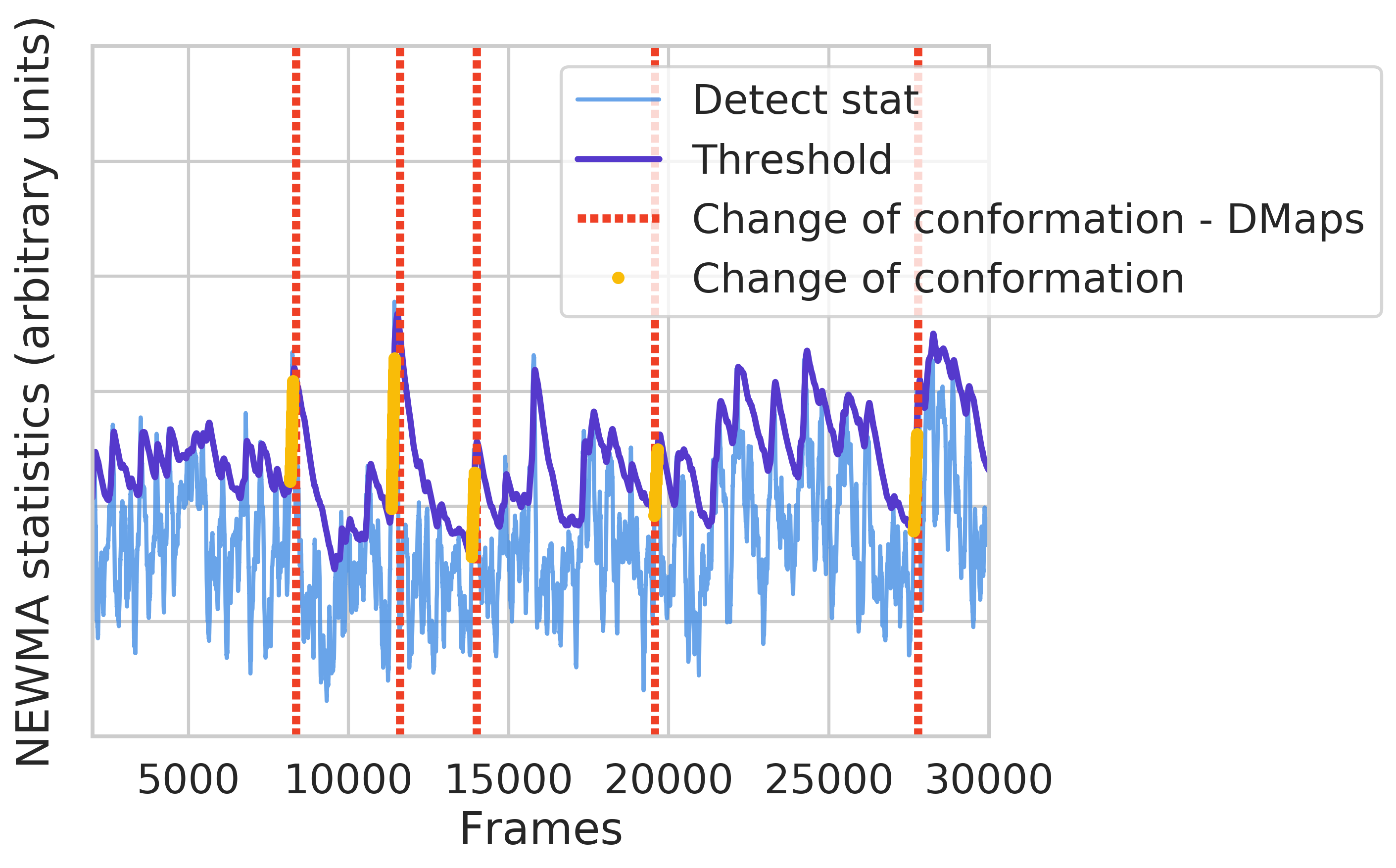}
\caption{NEWMA RP on OPU detection statistic (solid light blue line) and adaptive threshold (solid navy blue line) as a function of time frames, applied to the trajectory of the virus SARS-CoV-2. When the detection statistic is larger than the threshold for a significant number of steps, we flag those steps as changes in conformation (golden dots). The changes of conformation as found by the changes in the spectrum of the diffusion operator are indicated by the vertical dotted red lines.}
\label{fig:eigenvalfntime}
\end{figure}

However, the MD simulations published by Cespugli, Durmaz, Steinkellner, and Gruber \cite{Cespugli2020} of the virus SARS-CoV-2 do not include any ground truth. That is, we do not know when exactly conformational changes are happening. We compare in Figure \ref{fig:eigenvalfntime} the transitions detected using the NEWMA algorithm to the transitions detected using changes in the spectrum of the diffusion operators (DMaps). The detection statistics of NEWMA being extremely noisy for our datasets, for a point to be flagged as a change of conformation, the detection statistics has to be larger than the threshold for the next $20$ points.

The conformational changes detected by NEWMA match the ones obtained through changes of the DMaps spectrum. This is very promising: as pointed out before, this method does not require any prior knowledge to detect the changes. By using the change points detected by NEWMA, we can reduce the number of times that we have to compute the spectrum of the diffusion operator by a factor $4$ (see Figure \ref{fig:eigenvalfntime}). This is substantial gain, as the spectrum of the diffusion operator requires the computation of pairwise distances between all atoms in the simulation and also requires the diagonalisation of that matrix. As trajectories typically consider a much larger number of time frames, this effect is amplified.

\subsection{Comparison With Other Methods: Visualisations}

Since March 27th, D.E. Shaw Research (DESRES) has been releasing MD trajectories related to SARS-CoV-2 using the supercomputer Anton 2  \cite{DESRES2020}. Moreover, videos visualizing the simulated behaviors are also available, which gives us a comparison point for NEWMA. We focus on two $10-\mu$s-simulations of the trimeric SARS-CoV-2 spike glycoprotein (DESRES-ANTON-$10897136$ and -$10897850$). The first trajectory --- (a) --- was initiated in the closed state (PDB entry 6VXX), and remained stable. The second trajectory --- (b) --- was initiated in a partially opened state (PDB entry 6VYB) and showed a part of the protein detaching a bit and then detaching even more.

\begin{figure}[htpb]
	\centering
	 \begin{subfigure}[b]{0.455\textwidth}
    	\includegraphics[width=\textwidth]{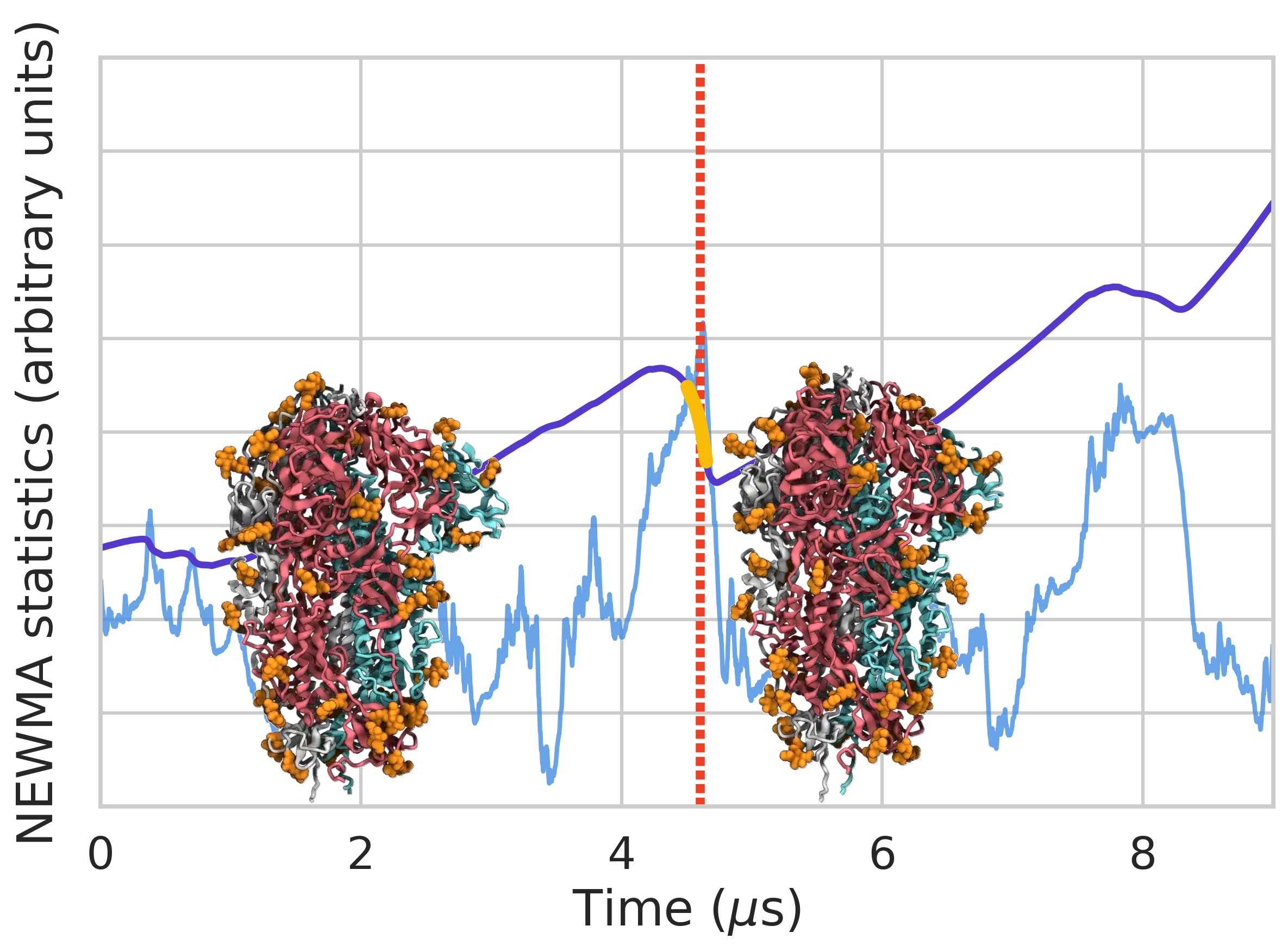}
    	\caption{Trajectory (a).}
    	\label{fig:desresnewmaclosed}
    \end{subfigure}%
    \begin{subfigure}[b]{0.545\textwidth}
    	\includegraphics[width=\textwidth]{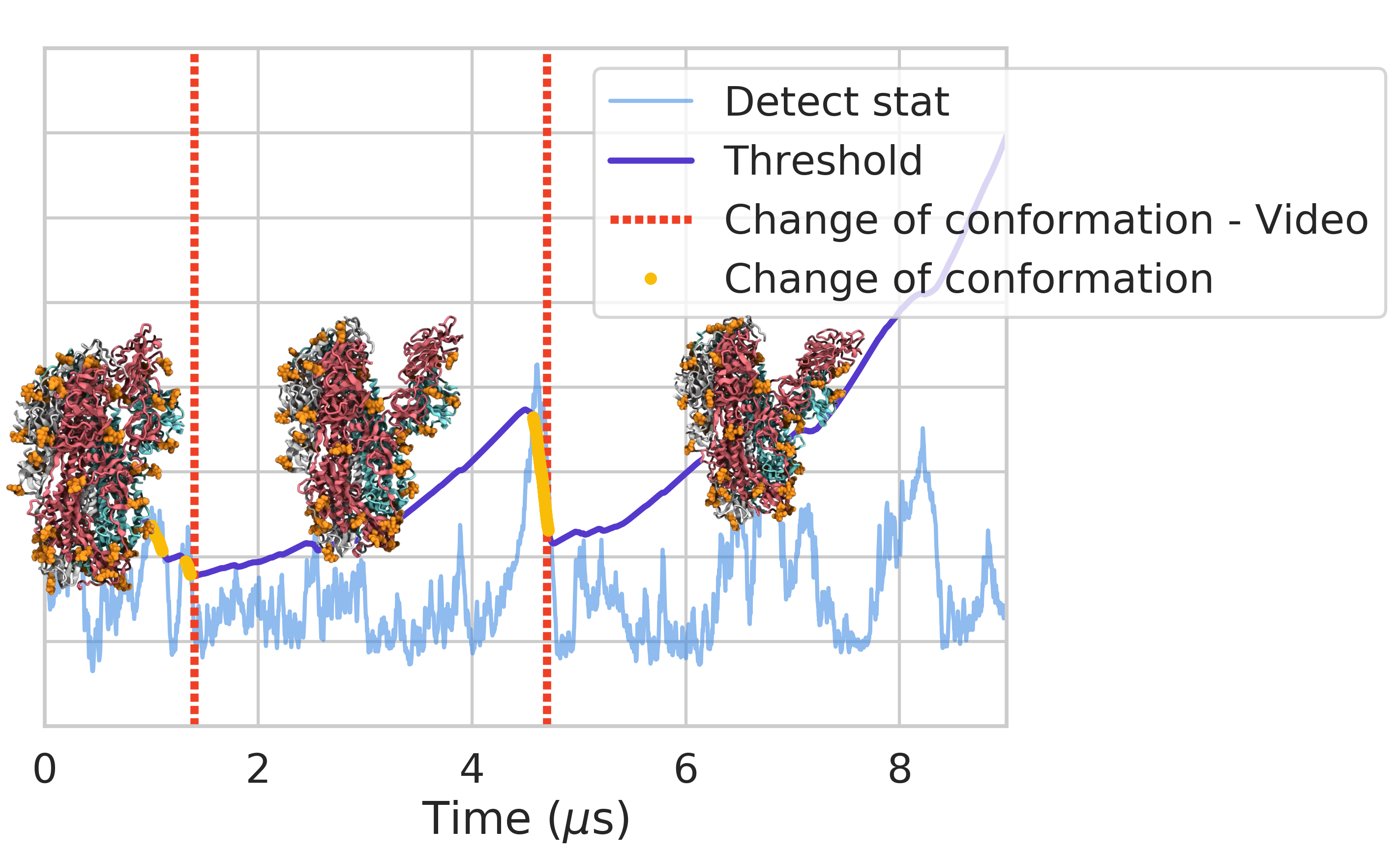}
    	\caption{Trajectory (b).}
    	\label{fig:desresnewmaopened}
    \end{subfigure}
	\caption{NEWMA RP on OPU detection statistic (solid light blue line) and adaptive threshold (solid navy blue line) as a function of time, applied to the trajectory (a) --- left pannel --- and (b) --- right pannel---. When the detection statistic is larger than the threshold, we flag those steps as changes of conformation (golden dots). The changes of conformation as observed in the video are shown by a vertical dotted red line.}
	\label{fig:desresnewma}
\end{figure}

The singular aspect of these computations is their sheer size: the glycoprotein counts no less than $n=46\ 851$ atoms, which amounts to $140\ 553$ features as each atom has three coordinates. With such scales, memory does become a problem if we consider using the algorithm NEWMA FF on CPU to compute RPs. This is not the case if optical random features (NEWMA RP on OPU) are used. Figure \ref{fig:desresnewma} shows the results obtained with NEWMA RP on OPU for trajectory (a) and (b, including images showing the changes observed in the videos.

We start by discussing trajectory (a) --- initiated in the closed state. While it is mostly stable, some ever-so-slight variations in the structure can be observed in its video visualisation --- in the upper-right and bottom-right parts of the glycoprotein. The change observed by NEWMA RP on OPU matches the one observed in the video --- see the representations included in Figure \ref{fig:desresnewmaclosed}.

Next, we discuss trajectory (b) --- initiated in a partially opened state. As mentioned before, this glycoprotein shows two major changes in its structure. First, the upper-right blob separates itself from the main structure. Then, it detaches further --- see the representations included in Figure \ref{fig:desresnewmaopened}. The observations of NEWMA RP on OPU are in agreement with those conformational changes.

\paragraph{Larger Scales}

While $140\ 553$ features may seem like a very large number, current computations in HPC for MD regularly goes far beyond that. In these regimes, it becomes extremely prohibitive to detect conformational changes on either a CPU or a GPU. DESRES also released all-inclusive versions of trajectories (a) and (b), that include water and ion atoms, in addition to the glycoprotein. All in all, this computation features more than $715\ 439$ atoms resulting in more than $2\ 146\ 317$ features. Note that while this may seem like a very large number, MD simulations sometimes include a couple of millions of atoms, or more. 

\begin{figure}
    \centering
    \includegraphics[width=.6\textwidth]{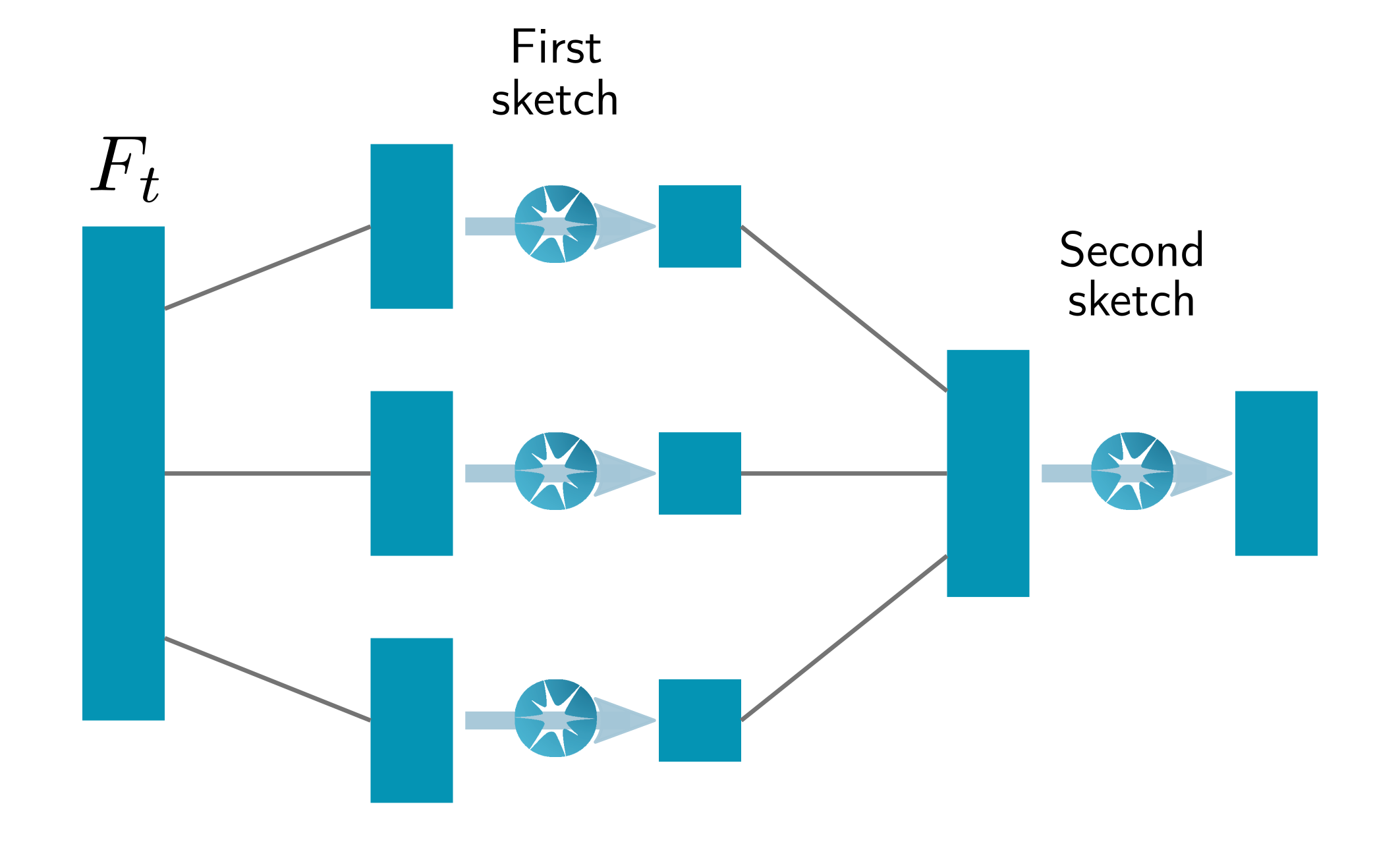}
    \caption{Pyramid sketch used to deal with high-dimensional data. The timeframe $F_t$, containing the three-dimensional coordinates of the atoms of the system, is divided to form three smaller vectors. Each of those vectors are projected using RPs into a lower-dimensional space, and the outputs are merged together. Finally, a second RP is used to project the data into a $k$-dimensional space.}
    \label{fig:pyramid}
\end{figure}

\begin{figure}[htpb]
	\centering
	\includegraphics[width=.6\textwidth]{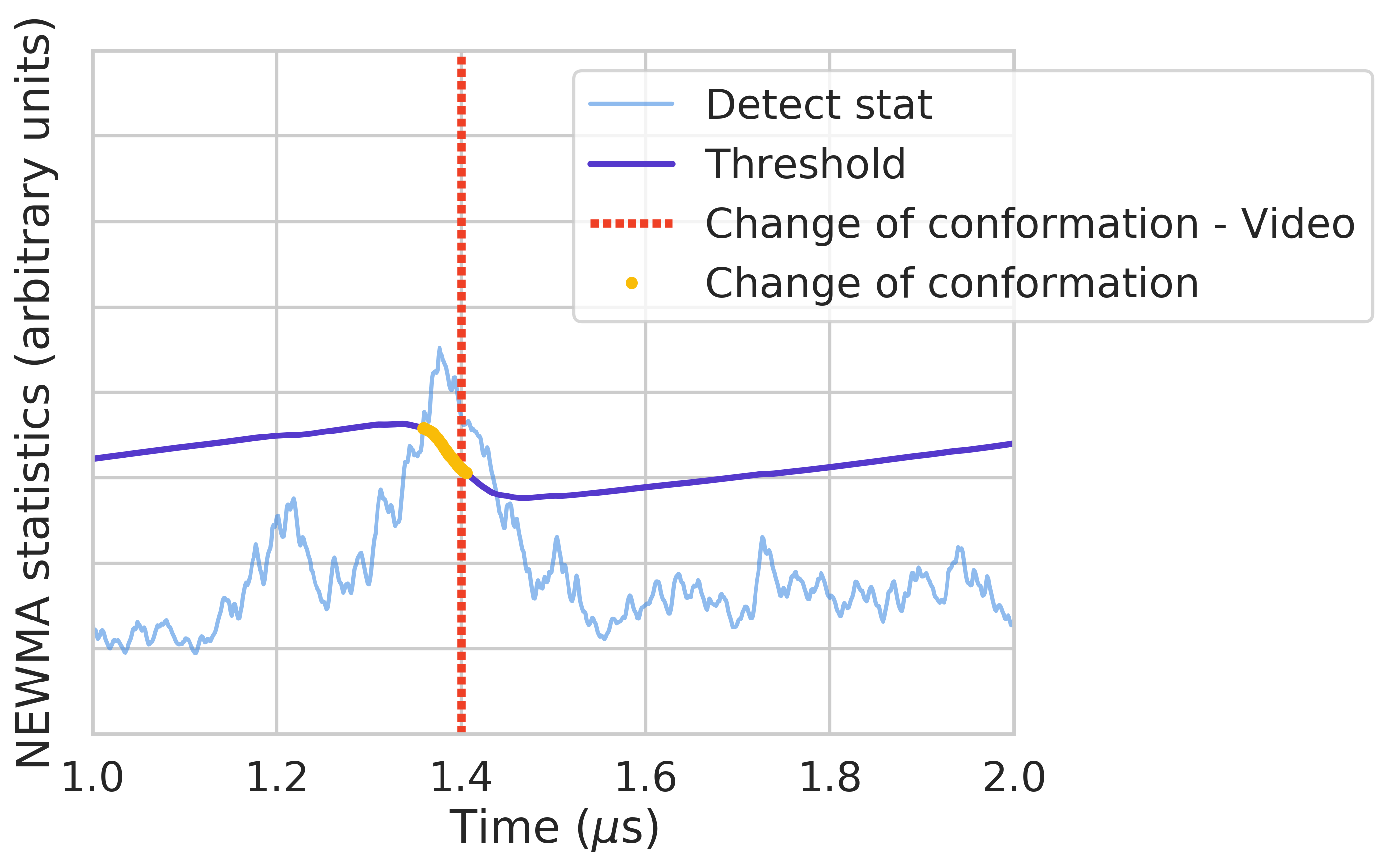}
	\caption{NEWMA RP on OPU detection statistic and adaptive threshold as a function of time, applied to the all-inclusive version of trajectory (b). We focus on the first few microseconds of the trajectory. Legend is the same as Figure \ref{fig:desresnewmaclosed}.}
	\label{fig:desresnewmaopenedFULL}
\end{figure}

Analysing such data on a CPU or a GPU would take too long and would require a very large amount of memory footprint. However, NEWMA RPs implemented on the OPU can be used to study the conformational changes of these all-inclusive versions. Note that a pyramidal sketch of the data is used to bypass the limit in the input size of the OPU --- see Figure \ref{fig:pyramid} for additional information. Figure \ref{fig:desresnewmaopenedFULL} shows the results obtained for the first microseconds of the all-inclusive version of (b): they match the behavior observed on the glycoprotein-atoms-only trajectory (b).

The experiments above show that the results obtained by applying NEWMA to MD simulations match those observed through the DMaps algorithm as well as changes observed in video visualisations. 

\subsection{Computational Cost: NEWMA on OPU or CPU}

\begin{figure}[htpb]
	\centering
	\includegraphics[width=.6\textwidth]{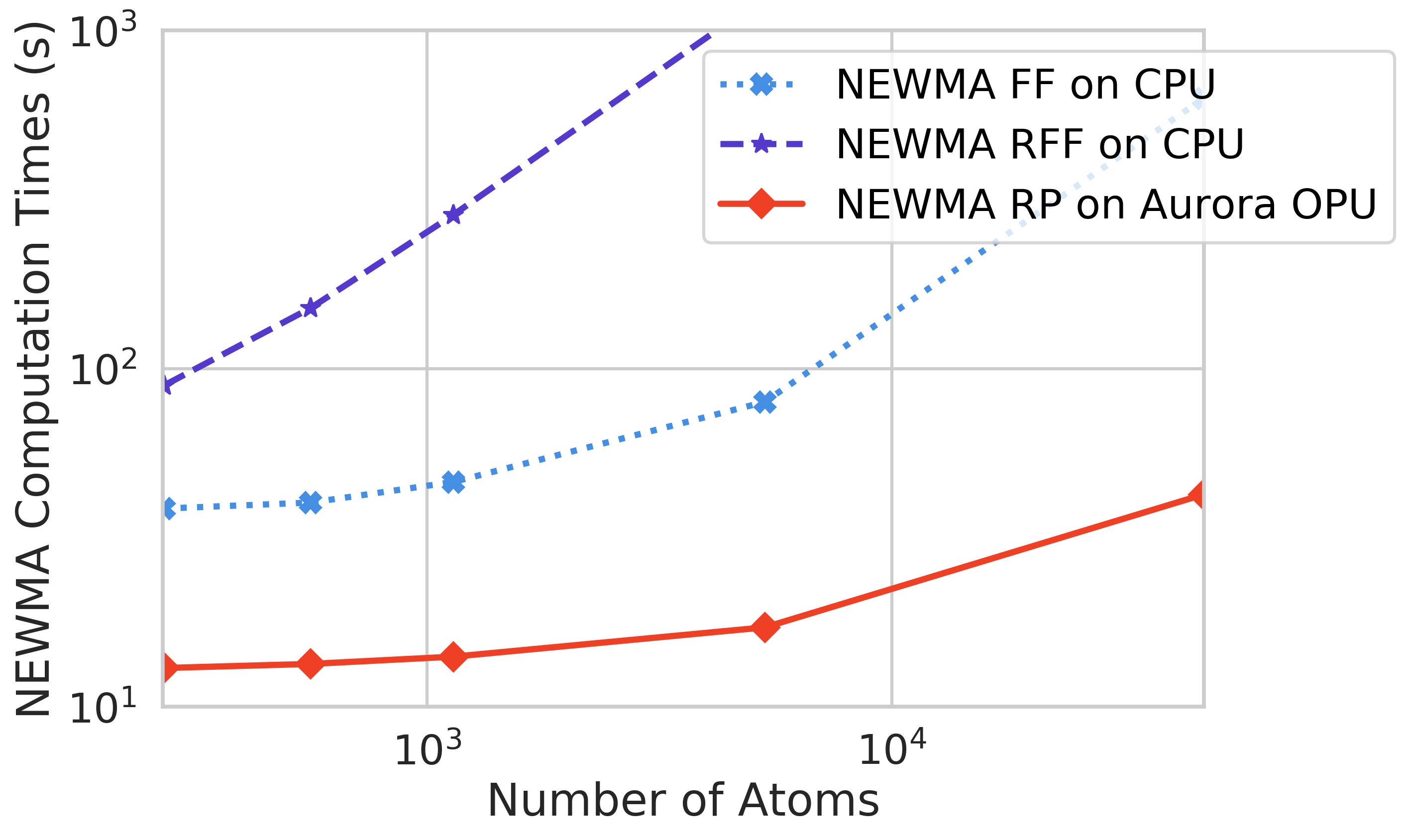}
	\caption{Computation times of NEWMA RP on Aurora OPU (solid red line and diamonds), NEWMA FF on CPU (dotted light blue line and crosses), and NEWMA RFF on CPU (dashed blue line and stars) as a function of the number of atoms of the various systems.}
	\label{fig:performance}
\end{figure}

We compare here the performances of NEWMA using the three methods presented in Section \ref{sec:newma} to compute RPs: NEWMA RFF on CPU, NEWMA FF on CPU, and NEWMA RP on OPU. In addition to the SARS-CoV-2 trajectories presented above \cite{Cespugli2020, DESRES2020}, we benchmarked NEWMA using supplementary MD simulations of molecular systems of various sizes provided by DESRES \cite{Lindorff2011}. We analyse $10\ 000$ times steps of all those trajectories. Figure \ref{fig:performance} shows the results of these investigations, comparing the computation times of the different implementations of NEWMA as a function of the number of atoms of the molecular system studied.

By design, NEWMA RFF on CPU is slower than NEWMA FF on CPU. However, it is interesting to note that NEWMA RP on OPU is always faster than both CPU-using methods. While the difference is not significant for a number of atoms under $\mathcal{O} \left( 10^3 \right)$, it increases as it grows larger. In fact, for trajectories (a) and (b), computation times were ten times faster on an OPU than on a CPU. 

More importantly, it still took less than a minute to perform those analyses on an OPU. By comparison, Anton 2, the supercomputers used to produce trajectory (a) and (b), can simulate around $80\ \mu$s a day for systems of these sizes \cite{Anton2014}. This translates to approximately $3$ hours to produce one of such trajectories, which can be analyzed using NEWMA in about $40$ s. 

Using optical random features particularly shines as the computation time is almost independent of the number of atoms. Indeed, by design, the NEWMA RP on OPU provides a scalable method to analyse conformational changes in MD simulations. While the all-inclusive version of trajectory (b) was not included in Figure \ref{fig:performance}, as we did not run the CPU-based NEWMA methods, it only took about two minutes to analyse it. This is a couple of orders of magnitude faster than the time it would take most people to download the $62$ GB of this trajectory.

\section{Conclusions and Outlooks}
\label{sec:conclusion}

We presented and implemented the use of NEWMA, a scalable, model-free change-point detection method, to identify conformational changes. In addition, we also proposed a strategy combining NEWMA and DMaps to learn CVs and enhance sampling in MD. 

We showed that using NEWMA is a very effective way to detect transitions in MD for large molecular systems. This way, the DMaps algorithm --- which requires to compute the diffusion matrix and extracting its spectrum --- need only to be applied once a change is detected by NEWMA, as a mean to learn CVs. This makes the combination of these two methods a powerful strategy to enhance sampling in MD. They provide a clever and efficient way to analyse MD trajectories online, enabling more resources to be allocated to the computation of said trajectories.

While CPU-based algorithms reach their limits with the number of features of these MD simulations, we showed that using a LightOn OPU bypasses this issue. Optical random features make the use of NEWMA virtually independent on the number of atoms in the trajectory, both in terms of memory and computational costs. In addition, as the size of the data grows, the OPU can be used to perform a pyramidal sketch. While the example presented above was made of two layers of RPs, the same idea could easily be extended, adding more layers for higher-dimensional data. This allows the seamless of use NEWMA on high dimensional data. This could be especially promising for systems such as lipid bilayers, which include billions of atoms. 

In addition, NEWMA is an online detection method, that can detect conformational changes just as they occur in the simulated trajectory, without the need to store the entire trajectory. In the context of MD simulations, using optical RPs with the OPU allows the full HPC infrastructure to solely focus on the actual computation of trajectories.

We believe that NEWMA RP on OPU could be of great help in many domains relying on MD, such as drug discovery.

\section*{Acknowledgements}

The authors acknowledge Igor Carron for motivating and driving this work. Thanks to Zofia Trstanova for helpful discussions, and to Caroline Chavier and the WiMLDS Paris team for sparking these conversations. In addition, we are grateful to the Iktos team, in particular to Nicolas Martin, for useful insights. Finally, we would like to thank D. E. Shaw Research for providing us molecular dynamics trajectories as well as supplementary information. 

\bibliographystyle{unsrt}
\bibliography{references}

\renewcommand{\appendixpagename}{Appendix}
\newpage
\appendix
\appendixpage
\setcounter{figure}{0} \renewcommand{\thefigure}{A.\arabic{figure}} 
\setcounter{table}{0} \renewcommand{\thetable}{A.\arabic{table}} 
\setcounter{equation}{0}
\renewcommand{\theequation}{A.\arabic{equation}} 

This Appendix provides additional information on the DMaps algorithm. Section \ref{sec:swissroll} showcases the example of the Swiss roll data set and introduces Pearson's correlation coefficients. The choice of DMaps hyperparameters is discussed in Section \ref{sec:hyperparam}.

\section{Numerical Example: Swiss Roll Data Set}
\label{sec:swissroll}

The Swiss roll data set was created as a means to test dimensionality reduction algorithms. In particular, linear techniques such as Principal Component Analysis (PCA) struggle to unravel its geometrical structure.

\begin{figure}[htpb]
    \centering
    \begin{subfigure}[b]{0.45\textwidth}
    	\includegraphics[width=\textwidth]{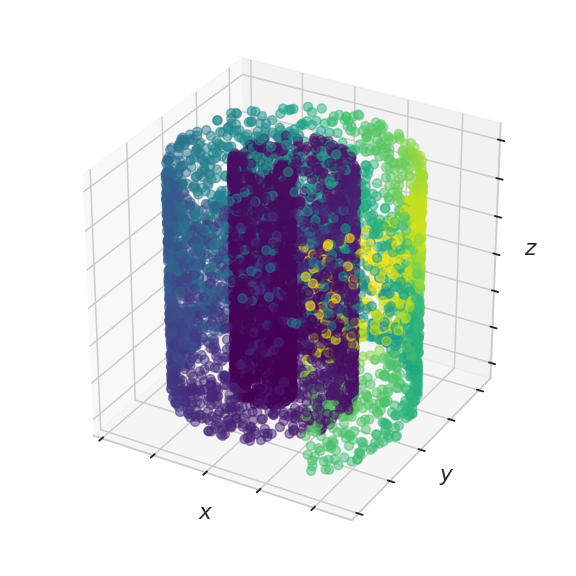}
    	\caption{Original space.}
    	\label{fig:swissrolloriginal}
    \end{subfigure}%
    \begin{subfigure}[b]{0.45\textwidth}
    	\includegraphics[width=\textwidth]{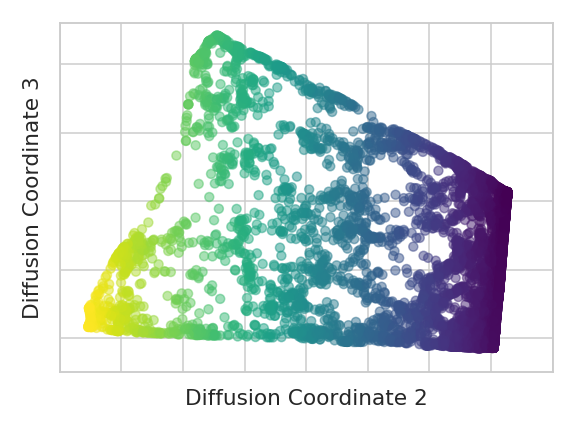}
    	\caption{Diffusion space.}
    	\label{fig:swissrolldm}
    \end{subfigure}
    \caption{The Swiss roll data set represented in the original three-dimensional space (Fig. \ref{fig:swissrolloriginal}) and in the two-dimensional space embedding given by the two dominant DCs (Fig. \ref{fig:swissrolldm}). The color map corresponds to the value of the second DC.}
    \label{fig:swiss_roll}
\end{figure}

To generate this data set, we sample $m=10\ 000$ points in a three-dimensional space, according to the general formula
\begin{equation}
\begin{array}{c}
x = \frac{1}{6} \left(\phi + \sigma \xi \right) \sin{\phi}, \\
y = \frac{1}{6} \left(\phi + \sigma \xi \right) \cos{\phi}, \\
z = \psi,
\end{array}
\end{equation}
where $\phi$ and $\psi$ are random numbers sampled uniformly from $\left[0,15\right]$. We add noise to our data by sampling uniformly a second random number, $\xi$, from $\left[0,1\right]$ and multiplying it by the noise strength $\sigma = 0.1$. Figure \ref{fig:swissrolloriginal} shows the distribution of points in three dimensions.

We apply the diffusion map algorithm, using the following parameters: 
\begin{itemize}
    \item $k=3$, since the data set is originally in three dimensions. Note that this comes down to representing the data points in a two-dimensional space as the first eigenvalue and eigenvector pair is discarded.
    \item $\alpha = 1/2$, which makes the DMaps operator coincide with the Fokker-Planck diffusion operator.
    \item $\gamma = 7$ and $t=1$, which have been chosen through trial and error.
\end{itemize}
 We discard the first pair of eigenvalue and eigenvector, and represent the points in a two-dimensional space parametrized by the other two DCs. The results are shown in Fig. \ref{fig:swissrolldm}. From these results, it appears that the data points can be easily embedded in a two-dimensional space: points that are close together on the roll are indeed close together in the embedded space. 

It is possible to compute the Pearson's correlation coefficients between the second and third DCs and the physical coordinates of the problem: $x$, $y$, $z$, $\phi$ and $\xi$. Pearson's correlation coefficients give us more information about the structure of the data by showing which of the original coordinates matter most in the geometrical description of the data set. They are defined as 
\begin{equation}
\label{eq:pearson}
\rho_{X, Y} = \frac{\mathrm{cov} \left( X, Y \right)}{\sigma_X \sigma_Y},
\end{equation}
where $X$ and $Y$ are two random one-dimensional variables of respective standard deviations $\sigma_X$ and $\sigma_Y$, and $\mathrm{cov}$ is the covariance. Such a coefficient ranges from $-1$ to $1$: a value of $1$ implies than $X$ and $Y$ are linearly related, with $X$ increasing as $Y$ increases, while a value of $-1$ means than $X$ decreases as $Y$ increases. A value of $0$ indicates that there is no linear relation between $X$ and $Y$. 

In the case of the Swiss roll data set, we find that the DC 2 is most correlated to $\phi$, with DC 2 decreasing as $\phi$ increases. The third DC is most correlated to $z$, with DC 3 decreasing as $z$ increases. This is what we expect from the way the data points were generated, and what can be observed in Fig. \ref{fig:swiss_roll}.

\section{Diffusion Maps Hyperparameters}
\label{sec:hyperparam}

The DMaps algorithm (Algorithm \ref{alg:DM}) depends on four hyperparameters: the width of the kernel $\gamma$, the normalization parameter $\alpha$, the time scale $t$, and the number of DCs $k$. In this work, we set $\alpha = 1/2$ so that the Markov chain described by the diffusion matrix $M$ approximates the Fokker-Planck diffusion process. We also set $t=1$ as we found the value of $t$ did not have much impact on the results observed.

It is tricky to find an appropriate value for the width of the kernel that unravels the geometrical structure of a given data set. Indeed, defining a cost function is impossible as we do not have a target to compare to. The inverse of the median of the pairwise distances between the points of the data set provides a good candidate for $\gamma$.  This ensures that, for at least half of our data points, the RBF kernel is larger than a certain value. We use this method, known as the \textit{median trick} \cite{scholkopf2001}, through this report. 

As for the number of relevant diffusion coordinates, our goal is to find the lowest value for the dimension of the diffusion space $k$ retaining the relevant features of our data set. Thus, we define a cost function with a regularization term which penalises model with a larger value of $k$. For the sake of simplicity, in this section, we set $t=1$ and remove the dependence on this parameter from our equations. 

The \textit{diffusion distance} between two vector points $x_i,\ x_j$ at time $t$ is defined so that it is the Euclidean distance in the diffusion embedding
\begin{equation}
D^k \left( x_i,\, x_j \right)^2 = \Vert \Psi^k \left(x_i \right)  -  \Psi^k \left(x_j \right)  \Vert^2.
\end{equation}

We define the error for the value $k$ as 
\begin{equation}
E_k = \sum_{i=1}^{m-1} \sum_{j=i+1}^{m} \left( D^k \left( x_i,\, x_j \right)^2 - D^m \left( x_i,\, x_j \right)^2 \right)^2,
\label{eq:err}
\end{equation}
that is, the difference between the diffusion distance computed using only the first $k$ components of the spectrum and the diffusion distance computed using the full spectrum of the diffusion matrix. Obviously, this function is strictly decreasing with $k$ and cannot be used as a cost function itself, as its minimum is reached for $k=m$. 

In order to build our cost function, we introduce the regularization term
\begin{equation}
R_k = \sum_{i=1}^{k} \frac{1}{\left( \lambda_i \Vert \psi_i \Vert_{\ell_1} \right)^\delta},
\label{eq:reg}
\end{equation}
with $\delta \in \mathbb{Z}^*$. This term penalises models which use a larger value of $k$, including eigenvalues which are very small. The introduction of such a regularization term stems from the fact that $\lambda_i \Vert \psi_i \Vert_{\ell_1}$ is a decreasing function of $i$, with values getting quickly suppressed and with an inflection point which seems to corresponds to the best fit value of $k$.

Our cost function is then the sum of our error and regularization terms 
\begin{equation}
J_k \left( \beta \right) = E_k + \beta R_k, 
\end{equation}
where $\beta \in \mathbb{R}^{+*}$ is a parameter to be determined. For the sake of simplicity, we introduce and represent 
\begin{equation}
F_k \left( \beta \right) = \frac{J_k \left( \beta \right)}{\beta},
\label{eq:costfunc}
\end{equation}
such that $\lim_{k\rightarrow +\infty}{F_k \left( \beta \right)} = \lim_{k\rightarrow +\infty}{R_k}$ is independent of the value of the value of $\beta$. Finding a proper value for $\beta$ is crucial, as we will discuss in the next subsection. From experiments, it appears that the value $\beta = E_{k=1}$ seems to provide good results on various data sets. 
\begin{figure}
    \centering
    \includegraphics[width=.5\textwidth]{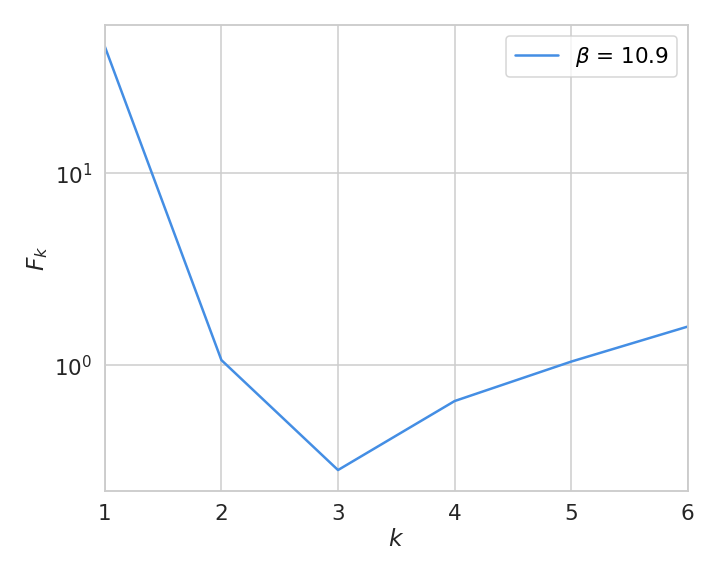}
    \caption{Cost function (Eq. \ref{eq:costfunc}) as a function of increasing values of $k$, for different values of $\beta$, for the Swiss roll data set.}
    \label{fig:swissrollcostfunc}
\end{figure}

Figure \ref{fig:swissrollcostfunc} shows the cost function \eqref{eq:costfunc} obtained with $\delta = 1$. As expected for the Swiss roll data set, it reaches a minimum for $k=3$. 

\end{document}